\documentclass[12pt]{article}
\usepackage[T1]{fontenc}
\usepackage[latin9]{inputenc}
\usepackage{verbatim}
\usepackage{float}
\usepackage{amsthm}
\usepackage{amsmath}
\usepackage{amssymb}
\usepackage{algorithmic}
\usepackage{algorithm}

\usepackage{subfigure}

\makeatletter

\floatstyle{ruled}

\numberwithin{equation}{section}
\numberwithin{figure}{section}
\theoremstyle{plain}
\newtheorem{thm}{\protect\theoremname}
  \theoremstyle{definition}
  \newtheorem{example}[thm]{\protect\examplename}
  \theoremstyle{definition}

\@ifundefined{definecolor}
 {\usepackage{color}}{}

\newtheorem{corollary}{Corollary}

\newtheorem{theorem}{Theorem}
\newtheorem{definition}{Definition}

\newtheorem{observation}{Observation}

\newcommand{\dahntab}[1]{
  \newbox\mybok%
  \setbox\mybok=\hbox{\vbox{
      \begin{tabbing}
        #1
      \end{tabbing}%
    }}

  \newdimen\bokwidth%
  \bokwidth=\wd\mybok%
  \newdimen\myl%
  \myl=\textwidth%
  \divide\myl by 2%
  \divide\bokwidth by -2%
  \advance\myl by\bokwidth%
  \vrule width\myl height 0pt depth 0pt%
  \usebox\mybok%
}

\usepackage{tikz}
\usepgflibrary{shapes.symbols} 
\usepgflibrary[shapes.symbols] 
\usetikzlibrary{shapes.symbols} 
\usetikzlibrary[shapes.symbols] 
\usepgflibrary{shapes.geometric} 
\usepgflibrary[shapes.geometric] 
\usetikzlibrary{shapes.geometric} 
\usetikzlibrary[shapes.geometric] 

\makeatother

  \providecommand{\definitionname}{Definition}
  \providecommand{\examplename}{Example}
\providecommand{\theoremname}{Theorem}

\begin{document}
\title{\textbf{\Large Improved approximation algorithms for low-density
instances of the Minimum Entropy Set Cover Problem.} }

\author{Cosmin Bonchi\c{s}\thanks{Department of Computer Science, West University of Timi\c{s}oara,
Bd. V. Pârvan 4, Timi\c{s}oara, RO-300223, Romania.}, Gabriel Istrate\thanks{e-Austria Research Institute, Bd. V. Pârvan 4, cam. 045
B, Timi\c{s}oara, RO-300223, Romania.email:gabrielistrate@acm.org. corresponding author}}

\maketitle

\begin{abstract}

We study the approximability of instances of the {\em minimum entropy set cover} problem, parameterized by 
the average frequency of a random element in the covering sets. We analyze an algorithm combining a greedy 
approach with another one biased towards large sets. The algorithm is controled by the percentage of 
elements to which we apply the biased approach. The optimal parameter choice has a {\em phase transition} 
around average density $e$ and leads to improved approximation guarantees when average element frequency 
is less than $e$. 
\end{abstract}

\section{Introduction}

The {\em minimum entropy set cover problem} (MESC) \cite{halperin-karp-sc} arose from a maximum likelihood 
approach to haplotype inference in computational biology (see also \cite{mandoiu2005haplotype}). Halperin 
and Karp showed that the problem is NP-complete and provided an additive upper bound (equal to three) on the performance of the Greedy algorithm. This was later improved by Cardinal et
 al.~\cite{tight-minentropy-setcover}, who showed a tight additive upper bound of $\log_{2}(e)$. 
Cardinal et al. \\ ~\cite{cardinal2009minimum}  also studied several versions of this problem, notably minimum 
entropy graph coloring \cite{cardinal2004minimum} and minimum entropy orientation \cite{cardinal2008minimum}, as well as a generalization to arbitrary objective functions \cite{cardinal2008pmean}. Minimum entropy graph coloring has found applications to problems related to functional compression in information theory \cite{cardinal2004minimum}.

Minimum entropy set cover also lies behind a recently proposed family of measures of worst-case fairness in cost allocations in cooperative game theory \cite{istrate-bonchis2012-tugames}. This was accomplished by first studying \cite{istrate-bonchis2011} a minimum entropy version of the well-known {\em submodular set cover problem} \cite{wolsey-submodular,fujita-ssc}. Submodularity corresponds in the setting of cooperative game theory to {\em concavity} of the associated game, a property that guarantees many useful features of the game such as the non-emptiness of the core, membership of the Shapley value in the core, equivalence between group-strategyproofness and cross-monotonicity in mechanism design \cite{moulin1999incremental} and so on. 

In this paper we further study MESC restricted to sparse instances, that is to instances of Set Cover parameterized by $f$ (formally defined below), the average number of sets that cover a random element. In the spirit of the minimum entropy orientation problem (a version of MESC for which $f=2$) we aim to provide better approximation guarantees than those valid for the Greedy algorithm. To accomplish this goal we study the performance of an approximation algorithm $BiasedGreedy(\delta)$ parameterized by a constant $\delta\in [0,1].$

Our main result can be summarized as follows: we give general upper bounds on the performance of our proposed algorithm. These bounds improve on the approximation guarantee of the greedy algorithm when average element frequency is less than the constant $e.$ Furthermore, the best choice of control parameter $\delta$ depends on this frequency: it corresponds to the choice of a "biased" algorithm below critical value $e,$ and to the greedy algorithm above it.
 
The paper is structured as follows: in Section 2 we review basic notions and define the algorithm BiasedGreedy. The main result is presented and further discussed in Section 3. Its proof is given in Section 4. Next we present several applications of our main result to the Minimum Entropy Graph Coloring problem.

\section{Preliminaries}

In this paper we need the definition of Shannon entropy and its associated {\em divergence} of two distributions $P$ and $Q$:
\[
D(P\parallel Q) = \sum_i p_i \log_2 \frac{p_i}{q_i}. 
\]

We recall that $D(P \parallel Q)\geq 0$ for all $P$ and $Q.$

~

We are concerned with the following problem: 

\begin{definition}{\bf [MINIMUM ENTROPY SET COVER (MESC)]:}
 Let $U=\{u_{1}, u_{2}, \ldots , u_{n}\}$ be an $n-$element ground set, for some $n\geq 1,$ and let $\mathcal{P}=\{P_{1},P_{2},\ldots, P_{m}\}$ be a family of subsets of $U$ which cover 
$U$. A {\em cover} is a function $g:U\rightarrow [m]$ such that for every $1\leq i \leq n$, 
\[
 u_{i}\in P_{g(u_i)}\mbox{(``$u_{i}$ is covered by set $P_{g(u_i)}$'')}
\]
The entropy of cover $g$ is defined by:
\begin{equation} 
Ent(g)=-\sum_{i=1}^{m} \frac{|g^{-1}(i)|}{|U|}\log_2\frac{|g^{-1}(i)|}{|U|}.
\label{entcover}
\end{equation} 
[OBJECTIVE:]  Find a cover $g$ of minimum entropy. 
\label{def-mesc}
\end{definition} 

Consider an instance $(U,\mathcal{P})$ as above. Define
\[
f = \frac{\sum_{i=1}^{m}|P_i|}{|U|},
\]
the {\em average frequency of a random element in $U$.}

In the algorithm below we divide the elements of the ground set into {\em Light} and {\em Heavy} elements, based on their frequency of occurrence. Parameter $\delta$ controls this division: the least frequent $\delta n$ elements are deemed {\em Light}, while the rest are considered {\em Heavy}. 

Informally, the algorithm will first covers Light elements in a {\em biased} manner, simultaneously covering each such elemen by a set of maximum cardinality containing it. Once this phase is complete all Light elements are deleted from all sets. The Heavy elements are handled in an incremental manner via a Greedy approach. The algorithm is formally presented in the following:

\begin{figure}
\begin{center}
\textbf{\small INPUT}{\small : An instance $(U,\mathcal{P})$ of MESC}\\
{\small ~}\\
{\small $\mathcal{P}^{H}:=\{P_{1}^{H},P_{2}^{H},\dots,P_{k}^{H}\}$
where $P_{i}^{H}=P_{i}\setminus L$ for all $i\in[k]$}\\
{\small ~}\\
{\small While (there exists $e\in L$) }{\small \par}
{\small ~~~~~ choose $i_{e}\in[k]$ to maximize $|P_{i_{e}}|$
where $P_{i_{e}}\ni e$;}{\small \par}
{\small ~~~~~ let $g(e)=i_{e}$;}{\small \par}
{\small ~~~~~ $L:=L\setminus\{e\}$; }{\small \par}
{\small ~}\\
{\small While (there exists $e\in H$) }{\small \par}
{\small ~~~~~ choose $i_{e}\in[k]$ to maximize $|P_{i_{e}}^{H}|$
where $P_{i_{e}}^{H}\ni e$;}{\small \par}
{\small ~~~~~ let $g(e)=i_{e}$;}{\small \par}
{\small ~~~~~ erase $e$ from all $P_{i}^{H};$}{\small \par}
{\small ~~~~~ $H:=H\setminus\{e\}$; }{\small \par}
\textbf{\small ~}\\
\textbf{\small OUTPUT}{\small : the cover $g.$}

\end{center}
\caption{BiasedGreedy($\delta$)}
\end{figure}

\section{Main result}

Our main result shows that the following upper bound on the performance of algorithm BiasedGreedy holds:

\begin{theorem}
Algorithm BiasedGreedy($\delta$) produces a cover $BG:U\longmapsto[k]$ satisfying:
\begin{equation}
Ent(BG) \leq Ent(OPT)-(1-\delta)\log_{2}\left(\frac{1-\delta}{e}\right)+\log_{2}f + o(1).
\label{main-ineq}
\end{equation}
\label{main-thm}
\end{theorem}

\begin{corollary} 
The Biased algorithm, defined as the BiasedGreedy algorithm with $\delta = 1$, produces a cover $BI$ whose entropy satisfies 
\begin{equation}
Ent(BI) \leq Ent(OPT)+\log_{2} f.
\label{cor-ineq}
\end{equation}
\end{corollary} 

\begin{observation}
Optimizing over constant $\delta$ in inequality~(\ref{main-ineq}) reveals an interesting fact: the optimal choice of $\delta$ is always $\delta\in \{0,1\}$, i.e. the pure Biased or Greedy algorithms. More precisely
\begin{itemize} 
\item choice $\delta = 1$ (i.e. Biased) is optimal for $f< e$. 
\item when $f> e$ choice $\delta = 0$ (i.e. Greedy) becomes best.  
\end{itemize}
Thus the optimal choice for $\delta$ has a {\em phase transition} from $\delta = 1$ to $\delta = 0$ around average density $f=e$. 
\end{observation} 

\section{Proof of the main result}
\begin{proof}
Let $BG$ be the cover generated by the BiasedGreedy algorithm, and denote by $p_{i}^{\flat}=\frac{|BG^{-1}(i)|}{n}$ the associated probability distribution.

If $OPT$ is the optimal solution of the same instance,
denote $x_{i}=\left|OPT^{-1}(i)\right|$ and $y_{i}=\left|OPT^{-1}(i)\cap Heavy\right|$ for all $1\leq i\leq k$. 
By choice of $\delta,$ $ \sum_{i=1}^{k}y_{i} =n-\lceil \delta n \rceil \leq (1-\delta)n$ while $ \sum_{i=1}^{k}x_{i} =n.$

We rewrite the entropy of BG as follows:
\begin{align*}
Ent(BG)= & -\sum_{i=1}^{k}p_{i}^{\flat}\log_{2}p_{i}^{\flat}=-\sum_{i=1}^{k}p_{i}^{\flat}\log_{2}\left(|P_{i}|\frac{p_{i}^{\flat}}{|P_{i}|}\right)
\end{align*}

Denoting by $\#=(\#_{i})$ the distribution $\#_{i}=\frac{|P_{i}|}{\sum_{j=1}^{k}|P_{j}|}$
we obtain:

\begin{align}
Ent(BG)= & -\sum_{i=1}^{k}p_{i}^{\flat}\log_{2}|P_{i}|-\sum_{i=1}^{k}p_{i}^{\flat}\log_{2}\frac{p_{i}^{\flat}}{\#_{i}}+\log_{2}\sum_{i=1}^{k}|P_{i}|\nonumber \\
= & -\sum_{i=1}^{k}p_{i}^{\flat}\log_{2}|P_{i}|-D(BG\parallel\#)+\log_{2}\sum_{i=1}^{k}|P_{i}|\label{eq:entbg}
\end{align}

Considering now just the first sum we obtain

\begin{align*}
-\sum_{i=1}^{k}p_{i}^{\flat}\log_{2}|P_{i}|= & -\sum_{i=1}^{k}\frac{|BG^{-1}(i)|}{n}\log_{2}|P_{i}|=-\frac{1}{n}\sum_{i=1}^{k}\sum_{v\in BG^{-1}(i)}\log_{2}|P_{BG(v)}|\\
= & -\frac{1}{n}\sum_{v\in U}\log_{2}|P_{BG(v)}|=-\frac{1}{n}\sum_{v\in U}\log_{2}a_{v}
\end{align*}

where $a_{v}$ is the size of the set assigned by $BiasedGreedy$ to cover $v.$ 

Continuing, we infer

\begin{align*}
-\sum_{i=1}^{k}p_{i}^{\flat}\log_{2}|P_{i}|= & -\frac{1}{n}\sum_{i=1}^{k}\sum_{v\in OPT^{-1}(i)}\log_{2}a_{v}=-\frac{1}{n}\sum_{i=1}^{k}\log_{2}\prod_{v\in OPT^{-1}(i)}a_{v}\\
= & -\frac{1}{n}\sum_{i=1}^{k}\log_{2}\prod_{v\in OPT^{-1}(i)\cap Light}a_{v}\cdot\prod_{v\in OPT^{-1}(i)\cap Heavy}a_{v}
\end{align*}

From the definition of the algorithm we conclude the following:
\begin{itemize}
\item for all $v\in OPT^{-1}(i)\cap Light,$ 
\begin{align*}
a_{v}=&|P_{BG(v)}|=\max_{j,P_{j}\ni v}|P_{j}|\geq|P_{OPT(v)}|\geq|OPT^{-1}(i)| = x_i 
\end{align*}
\item
On the other hand, for $v\in OPT^{-1}(i)\cap Heavy$ we analyze the Greedy phase of BiasedGreedy algorithm in a manner completely similar to the analysis of the Greedy algorithm in \cite{tight-minentropy-setcover} and infer that  
\[
\prod_{v\in OPT^{-1}(i)\cap Heavy}a_{v}\geq y_{i}!
\] 
\end{itemize}
Therefore,
\begin{align*}
&-\sum_{i=1}^{k}p_{i}^{\flat}\log_{2}|P_{i}|\leq -\frac{1}{n}\sum_{i=1}^{k}\log_{2}\left(\prod_{v\in OPT^{-1}(i)\cap Light}x_{i}\right)\left(y_{i}!\right) =\\
&= -\frac{1}{n}\sum_{i=1}^{k}\log_{2}\left(x_{i}^{x_{i}-y_{i}}\right)\left(y_{i}!\right) = -\frac{1}{n}\sum_{i=1}^{k}(x_{i}-y_{i})\log_{2}x_{i}-\frac{1}{n}\sum_{i=1}^{k}\log_{2}y_{i}!\\
&= -\sum_{i=1}^{k}\frac{x_{i}}{n}\log_{2}\frac{x_{i}}{n}-\log_{2}n+\frac{1}{n}\sum_{i=1}^{k}y_{i}\log_{2}x_{i}-\frac{1}{n}\sum_{i=1}^{k}\log_{2}y_{i}!
\end{align*}
Applying now the inequality $y!\geq(y/e)^{y}$ we obtain:

\begin{align}
&-\sum_{i=1}^{k}p_{i}^{\flat}\log_{2}|P_{i}|\leq Ent(OPT)-\log_{2}n+\frac{1}{n}\sum_{i=1}^{k}y_{i}\log_{2}x_{i}-\frac{1}{n}\sum_{i=1}^{k}\log_{2}\frac{y_{i}^{_{y_{i}}}}{e^{y_{i}}} \nonumber\\
&= Ent(OPT)-\log_{2}n+\frac{1}{n}\sum_{i=1}^{k}y_{i}\log_{2}x_{i}-\frac{1}{n}\sum_{i=1}^{k}y_{i}\log_{2}y_{i}+\frac{1}{n}\sum_{i=1}^{k}y_{i}\log_{2}e \nonumber\\
&\leq Ent(OPT)-\log_{2}n-\frac{1}{n}\sum_{i=1}^{k}y_{i}\log_{2}\frac{y_{i}}{x_{i}}+(1-\delta)\log_{2}e \label{eq:entbg2}
\end{align}

Considering now distributions $\overline{x_{i}}=\frac{x_{i}}{n}$,
$\overline{y_{i}}=\frac{y_{i}}{\sum_j y_j}$ we obtain 

\begin{align*}
-\frac{1}{n}\sum_{i=1}^{k}y_{i}\log_{2}\frac{y_{i}}{x_{i}}= & -\frac{1}{n}\sum_{i=1}^{k}(n-\left\lceil \delta n\right\rceil )\overline{y_{i}}\log_{2}\frac{(n-\left\lceil \delta n\right\rceil )\overline{y_{i}}}{n\overline{x_{i}}}\\
= & -\frac{(n-\left\lceil \delta n\right\rceil )}{n}\sum_{i=1}^{k}\overline{y_{i}}\log_{2}\frac{\overline{y_{i}}}{\overline{x_{i}}}-\frac{(n-\left\lceil \delta n\right\rceil )}{n}\log_{2}\frac{(n-\left\lceil \delta n\right\rceil )}{n}\\
= & -\frac{(n-\left\lceil \delta n\right\rceil )}{n}D(\overline{y}\parallel\overline{x})-\frac{(n-\left\lceil \delta n\right\rceil )}{n}\log_{2}\frac{(n-\left\lceil \delta n\right\rceil )}{n}
\end{align*}

Putting all things together:

\begin{align*}
Ent(BG) \leq & Ent(OPT)-(1-\delta)\log_{2}(1-\delta)+(1-\delta)\log_{2}e+\log_{2} f + o(1).
\end{align*} 
and the proof is complete.\end{proof}

\section{Application to minimum entropy graph coloring}

Just as it is the case with the Greedy algorithm \cite{cardinal2009minimum}, our result has implications for the minimum entropy coloring problem. This problem can be recast as an {\em implicit set cover problem} \cite{Karp2011122}, where the sets are the maximal independent sets in $G$. Given the intractability of the maximum independent set problem, we can only efficiently implement the Biased algorithm on special classes of graphs, where this problem is easier. On the other hand algorithm Biased has some nice properties, similar to those discussed in \cite{cardinal2009minimum}) for the Greedy algorithm: 
\begin{enumerate} 
\item it can be implemented in polynomial time on perfect graphs. Indeed, the largest independent set containing a given vertex can easily be computed in a perfect graph. 
\item it allows the use of $\eta$-approximately optimal independent sets (for some constant $\eta\geq 1$) instead of optimal ones at the expense of introducing an extra factor of $\log_{2}(\eta)$ in the upper bound of equation~(\ref{main-ineq}). This follows easily by simply redoing the proof of the main Theorem in this setting. We can apply this observation to get a slight improvement of Theorem 8 from \cite{cardinal2009minimum} when $f<e$: 
\begin{corollary} 
Algorithm Biased produces a coloring of a graph $G=(V,E)$ with maximum degree $\Delta$ satisfying 
\[
Ent(Biased)\leq Ent(OPT)+\log_{2}(\Delta+2)+\log_{2}(f/3).
\]
\end{corollary}

The proof of the corollary directly parallels that of Theorem 8 from \cite{cardinal2009minimum}.

\end{enumerate} 

Applying Theorem~\ref{main-thm} to graph coloring problems is rather inconvenient as parameter $f$ involves maximal independent sets and is not easy to compute. The situation is slightly better for graphs with independence number $\alpha(G)\leq 3$. In this case maximal independent sets correspond either to triangles, edges, or isolated vertices in the complement graph $\overline{G}$. Parameter $f$ also has an easier interpretation: 
 Let $I$ be the number of isolated vertices in $\overline{G}$. Let $T$ be the number of distinct {\em triangles} in $\overline{G}$. Finally, let $M$ be the number of edges that are not contained in any triangle. Then 
\[
f=\frac{I+2M+3T}{n}
\]
Furthermore, in this case the algorithm Biased has a very natural interpretation: we create a tentative color $c_W$ for any maximal independent set (triangle, edge or isolated vertex) $W$ and add color $c_{W}$ to a list coloring of all vertices in $W$. Then  for each vertex we select a random color from its list. 

The algorithm Biased can be improved in practice by employing a number of heuristics such as: 
\begin{itemize} 
\item Attempt to color all elements of a largest independent set with the same color.  
\item Collapse two colors into one if legal. 
\end{itemize} 
These heuristics can only decrease the entropy of the resulting coloring.

There are instances (e.g. edge orientations of a cycle from \cite{cardinal2008minimum}) where Biased outperforms Greedy. But even when it doesn't, our analysis may provide better theoretical guarantees than those available for Greedy. 

\begin{example} {\em
Consider graph $G=(V,E)$ from Figure~\ref{graph-to-color} (a) (its complement is displayed in Figure~\ref{graph-to-color} (b)).

\begin{figure}[h] 
\begin{center} 
\begin{minipage}[t]{0.90\columnwidth}
  \begin{minipage}[t]{0.45\textwidth}%
    \begin{tikzpicture} 
      [
	elips/.style={shape=ellipse, inner ysep=1.5cm, inner xsep=2.0cm}, 
	vertex/.style={circle,fill=black!25,minimum size=17pt,inner sep=0pt} 
      ] \node[name=E,elips]{};
	\matrix [column sep={0.5cm,between origins}, row sep={0.5cm,between origins}, ] at (E.mid) {  
	  & &  \node[vertex] (n1) {2}; & & &  \node[vertex](n2) {3}; & & \\  
	  & & & & & & & \\  
	  \node[vertex] (n3) {1}; & & & & & & & \node[vertex] (n4) {4};  \\  
	  & & & & & & & \\  
	  \node[vertex] (n6) {8}; & & & & & & & \node[vertex] (n5) {5}; \\  
	  & & & & & & & \\  
	  & &  \node[vertex] (n7) {7}; & & & \node[vertex] (n8) {6}; & & \\ 
	};

      \draw (n1) -- (n1); 
      \draw (n1) -- (n4); 
      \draw (n1) -- (n5); 
      \draw (n1) -- (n6); 
      \draw (n1) -- (n7); 
      \draw (n1) -- (n8); 
      \draw (n2) -- (n2); 
      \draw (n2) -- (n5); 
      \draw (n2) -- (n6); 
      \draw (n2) -- (n7); 
      \draw (n3) -- (n3); 
      \draw (n3) -- (n4); 
      \draw (n3) -- (n5); 
      \draw (n3) -- (n6); 
      \draw (n3) -- (n7); 
      \draw (n3) -- (n8); 
      \draw (n4) -- (n4); 
      \draw (n4) -- (n6); 
      \draw (n4) -- (n7); 
      \draw (n5) -- (n5); 
      \draw (n5) -- (n6); 
      \draw (n5) -- (n7); 
      \draw (n5) -- (n8); 
      \draw (n6) -- (n6); 
      \draw (n7) -- (n7); 
      \draw (n8) -- (n8);
    \end{tikzpicture}
  \end{minipage}
~
  \begin{minipage}[t]{0.45\textwidth}
    \begin{tikzpicture} [
	elips/.style={shape=ellipse, inner ysep=1.5cm, inner xsep=2.0cm}, 
	vertex/.style={circle,fill=black!25,minimum size=17pt,inner sep=0pt} 
      ] 
      \node[name=E,elips]{};
      \matrix [column sep={0.5cm,between origins}, row sep={0.5cm,between origins}, ] at (E.mid) {  
	& &  \node[vertex] (n1) {2}; & & &  \node[vertex](n2) {3}; & & \\  
	& & & & & & & \\  
	\node[vertex] (n3) {1}; & & & & & & & \node[vertex] (n4) {4};  \\  
	& & & & & & & \\  
	\node[vertex] (n6) {8}; & & & & & & & \node[vertex] (n5) {5}; \\  
	& & & & & & & \\  
	& &  \node[vertex] (n7) {7}; & & & \node[vertex] (n8) {6}; & & \\ 
      };
      \foreach \i/\j in {1/2, 1/3, 3/2, 6/7, 7/8, 8/6, 8/2, 8/4, 2/4, 4/5}     
	\draw (n\i) -- (n\j);
    \end{tikzpicture} 
  \end{minipage}
\end{minipage} 
\caption{(a) graph $G$ (b) its complement $\overline{G}$}
\label{graph-to-color} 
\end{center} 
\end{figure}

\begin{figure}[h]
\begin{center} 
\begin{minipage}[t]{0.90\columnwidth}
\begin{minipage}[t]{.45\textwidth}
    \begin{center}
    \begin{tikzpicture} [
      elips/.style={shape=ellipse, inner ysep=1.5cm, inner xsep=2.0cm}, 
      vertex/.style={circle,draw=black,fill=black!5,minimum size=17pt,inner sep=0pt},
      v2/.style={circle,fill=black!25,minimum size=17pt,inner sep=0pt}, 
      v3/.style={circle,fill=black!55,minimum size=17pt,inner sep=0pt} 
      ] \node[name=E,elips]{};
      \matrix [column sep={0.5cm,between origins}, row sep={0.5cm,between origins}, ] at (E.mid) {  
      & &  \node[vertex] (n1) {a}; & & &  \node[vertex](n2) {a}; & & \\  
      & & & & & & & \\  
      \node[vertex] (n3) {a}; & & & & & & & \node[v3] (n4) {c};  \\  
      & & & & & & & \\  
      \node[v2] (n6) {b}; & & & & & & & \node[v3] (n5) {c}; \\  
      & & & & & & & \\  
      & &  \node[v2] (n7) {b}; & & & \node[v2] (n8) {b}; & & \\ 
      };
      \foreach \i/\j in {1/2, 1/3, 3/2, 6/7, 7/8, 8/6, 8/2, 8/4, 2/4, 4/5}     
	      \draw (n\i) -- (n\j);
    \end{tikzpicture} 
    \end{center}
  \end{minipage}
  \hfill
  \begin{minipage}[t]{.45\textwidth}
    \begin{center}
      \begin{tikzpicture} [
      elips/.style={shape=ellipse, inner ysep=1.5cm, inner xsep=2.0cm}, 
      vertex/.style={circle,draw=black,fill=black!2,minimum size=17pt,inner sep=0pt},
      v2/.style={circle,fill=black!15,minimum size=17pt,inner sep=0pt}, 
      v3/.style={circle,fill=black!30,minimum size=17pt,inner sep=0pt},
      v4/.style={circle,fill=black!55,minimum size=17pt,inner sep=0pt} 
      ] \node[name=E,elips]{};
      \matrix [column sep={0.5cm,between origins}, row sep={0.5cm,between origins}, ] at (E.mid) {  
      & &  \node[v2] (n1) {b}; & & &  \node[vertex](n2) {a}; & & \\  
      & & & & & & & \\  
      \node[v2] (n3) {b}; & & & & & & & \node[vertex] (n4) {a};  \\  
      & & & & & & & \\  
      \node[v3] (n6) {c}; & & & & & & & \node[v4] (n5) {d}; \\  
      & & & & & & & \\  
      & &  \node[v3] (n7) {c}; & & & \node[vertex] (n8) {a}; & & \\ 
      };
      \foreach \i/\j in {1/2, 1/3, 3/2, 6/7, 7/8, 8/6, 8/2, 8/4, 2/4, 4/5}     
	      \draw (n\i) -- (n\j);
      \end{tikzpicture} 
      \end{center}
  \end{minipage}
\end{minipage}
\end{center}
\label{c1c2}
\caption{Two colorings $C_{1}$ and $C_{2}$ of graph $G$. For convenience the complement graph $\overline{G}$ is pictured, rather than $G$. $C_{1}$ is an optimal solution.}
\end{figure}
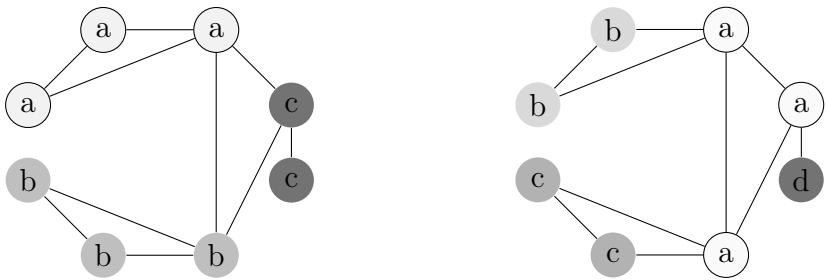

Graph $G$ provides an easy instance where Greedy and Biased (may) produce different colorings. Indeed, node 4 is colored by Biased with a color corresponding to a triangle, whereas 5 takes a color corresponding to an edge, so nodes 4 and 5 {\bf must} assume different colors in a Biased coloring, whereas they may have the same color in a Greedy coloring. With the optimizations described above both Greedy and Biased produce one of the two  colorings $C_{1},C_{2}$ from Figure~\ref{c1c2}, with color classes of cardinalities $(3;3;2;0)$ and $(3;2;2;1)$, respectively. The first one corresponds to the optimal solution. On the other hand for this graph the average element frequency $f=\frac{3\times 3+1\times 2}{8}=\frac{11}{8}<e$, so the upper bound on the entropy of coloring $C_{2}$ given by Corollary~\ref{cor-ineq} is tighter than the one provided by the Greedy algorithm in \cite{cardinal2009minimum}.
}
\end{example}

\section*{Acknowledgments}

Both authors contributed in a substantially equal manner to this work:
G.I. suggested the problem, performed research and wrote the paper.
C.B. performed research and wrote the paper.

The first author has been supported by a project on Postdoctoral national
programs, contract CNCSIS PD\_575/2010.

The corresponding author has been supported by CNCS IDEI Grant PN-II-ID-PCE-2011-3-0981
\textquotedbl{}Structure and computational difficulty in combinatorial
optimization: an interdisciplinary approach\textquotedbl{}.


\end{document}